\pdfoutput=0
\documentclass[11pt,a4paper]{article}

\usepackage{jcappub} 
\usepackage[T1]{fontenc} % if needed
\usepackage[utf8]{inputenc}
\usepackage{pstool}
\usepackage{hyperref}
\usepackage{amsmath,amssymb}
\usepackage{float}
\usepackage{graphicx}
\usepackage{graphics}
\usepackage{bm}
\usepackage{latexsym}
\usepackage{psfrag}
\usepackage{color}
\usepackage{subfigure}
\usepackage[normalem]{ulem}
\usepackage{textcomp}
\usepackage{comment}
\usepackage{mathalpha}
\usepackage{tabularx}
\usepackage{placeins}

\begin{document}

%title
\title{Weighing massive neutrinos with Lyman-$\alpha$ observations}

%author_list
\author{Anjan Kumar Sarkar$^{2\,\dagger}$ and}
\emailAdd{$^\dagger$asarkar@ncra.tifr.res.in}
\author{Shiv K. Sethi$^{1\,\ddagger}$}
\emailAdd{$^\ddagger$sethi@rri.res.in}

%author_affiliation
\affiliation{$^1$Raman Research Institute, C.~V.~Raman Avenue, Sadashivnagar, 
Bengaluru~560080, India}
\affiliation{$^2$National Centre for Radio Astrophysics, TIFR, Pune University 
Campus, Post Bag 3, Pune 411 007, India.}

%abstract
\abstract{ The presence of massive neutrinos has still not been revealed 
by the cosmological data. We consider 
a novel method based on the two-point line-of-sight correlation function of 
high-resolution Lyman-$\alpha$ data to achieve this end in the paper. 
We adopt semi-analytic models of Lyman-$\alpha$ 
clouds for the study. We employ Fisher matrix technique to show that 
it is possible to achieve a scenario in which
the covariance of the two-point function nearly vanishes for both the spectroscopic noise and the signal.
We analyze this near 'zero noise' outcome in detail   to argue it might be possible to detect neutrinos 
of mass range $m_\nu \simeq 0.05 \hbox{--}0.1 \, \rm eV$ with signal-to-noise of  unity with 
a single QSO line of sight.  We show that this estimate can be improved to SNR $\simeq 3\hbox{--}6$ with data  along  multiple line of sights
within the redshift range $z \simeq 2 \hbox{--} 2.5$. 
Such data sets already exist in the literature. 
We further carry out 
principal component analysis of the Fisher matrix to study the degeneracies
of the neutrino mass with other 
parameters. We show that Planck priors lift the degeneracies between
the neutrino mass and  other cosmological parameters.  
However, the prospects of the detection  of neutrino mass are driven by the  poorly-determined parameters characterizing the  ionization and  thermal state of Lyman-$\alpha$ clouds. We have also mentioned the possible limitations and 
observational challenges posed in measuring the neutrino mass using our method.  }

\maketitle

\section{Introduction}
\label{sec:intro}

In the past few decades, neutrino oscillation experiments have established 
that standard model neutrinos are massive. In particular, these experiments
measure mass differences between  the three neutrinos and  detect two eigenstates of non-zero masses: 
$\Delta m^2_{32} = (2.43 \pm 0.13) \times 10^{-3} \, {\rm eV}^2$ 
and $\Delta m^2_{21} = (7.59 \pm 0.21) \times 10^{-5} \, {\rm eV}^2$ (for details see e.g. \cite{SNO2004,adamson2008,KamLAND,athar2022status}).
The smaller mass difference is between electron and muon neutrinos and the larger mass difference 
is between the tau and either electron or muon neutrinos. These two mass differences do not allow
one to determine the hierarchy  of masses. If tau neutrino is more massive (Normal hierarchy), 
the sum of the three masses $\sum_i m_\nu^i > 0.045 \, \rm eV$. If electron/muon neutrinos are more massive (Inverted hierarchy),
then $\sum_i m_\nu^i > 0.09 \, \rm eV$.

It is well known that massive neutrinos have important implications for cosmology----light, massive neutrinos
constitute hot dark matter (HDM) in the universe. Unlike the particle physics data which only provides a lower limit on the sum of masses  the cosmological observables are sensitive to the total mass. From CMB and galaxy clustering data it is clear
that the dominant component of the dark matter is cold dark matter (CDM)  (e.g. \cite{aghanim2020} and references therein). 
Massive neutrinos provide a sub-dominant HDM component to the matter content of the universe. In the past decade, a number of studies have placed
constraints on the total neutrino mass $\sum m_{\nu}$ using different observational 
probes. Planck CMB temperature and polarization data
along with CMB lensing reconstruction yields the following constraint the HDM component: $\sum_i m_\nu^i < 0.27 \, \rm eV$. 
The constraint improves with the inclusion of low-redshift BAO data: $\sum_i m_\nu^i < 0.12 \, \rm eV$ (\cite{aghanim2020}). 
A further improvement, $\sum_i m_\nu^i < 0.09\, \rm eV$, has been reported following the inclusion of  the  redshift-space distortion analyses of the SDSS-IV eBOSS survey \cite{di2021}. ( All the  upper limits quoted above 
are at  95\% confidence level.)

 The CMB and low-redshift galaxy clustering  data are sensitive to scales $k \lesssim 0.2 \, \rm Mpc^{-1}$. Lyman-$\alpha$ data
 can potentially probe Jeans' scales in intergalactic medium in the redshift range $2.0 < z < 6$: $k \lesssim 5 \, \rm Mpc^{-1}$. 
 However, the analysis of Lyman-$\alpha$ data yield weaker bounds $\sum m_\nu^i < 0.5 \, \rm eV$ (e.g. \cite{yeche2017,palanque2020hints}
 use both high-spectral resolution Lyman-$\alpha$ data from XQ-100 and  HIRES/MIKE and low-resolution data from BOSS) largely 
 owing to uncertainty in thermal and dynamical state of Lyman-$\alpha$ clouds, which is degenerate with cosmological parameters (for a detailed
 discuss see e.g. \cite{palanque2020hints}). The joint analysis of Lyman-$\alpha$ data with Planck CMB and galaxy data significantly improves the 
 constraint: $\sum_i m_\nu^i < 0.09\, \rm eV$ \cite{palanque2020hints}. From this discussion it is clear that no 
 single cosmological probe can give requisite constraints on the sum of neutrino masses. While CMB anisotropies and 
 large-scale structure observations probe similar scales, a combination of the two is needed. CMB anisotropy measurement is 
 based on angular projection of linear scales (it is sensitive to $k\simeq \ell/\eta_0$ where $\eta_0=\int_0^{t_0}dt/a$ is the conformal
 time at the present) which  washes out information of perturbations at smaller scales. This situation is partly alleviated by
 large-scale clustering (e.g. BAO) data, but this data needs supplementary information from CMB data (e.g. prior information on parameters such
 as $n_s$ and $\Omega_m h^2$ which are determined very precisely by the CMB data).   The Lyman-$\alpha$ data is sensitive to 
 much smaller scales and hence it is potentially a better probe but it also needs precise 
 information on cosmological parameters  from CMB and galaxy data to be an effective probe of neutrino masses (e.g. \cite{pedersen2021neutrino,rossi2014suite,palanque2015neutrino,gratton2008prospects,gerbino2018status}). 

 The current study focuses on Lyman-$\alpha$ data as a probe of massive neutrinos. Our approach relies on semi-analytic
 modelling of Lyman-$\alpha$ clouds.  Lyman-$\alpha$ 
observations measure the contrast of  background QSO fluxes as a  function of the observing frequency. Our statistical estimator 
correspond to the 
two-point correlation function of the flux contrast in  frequency (redshift) space.   This function captures information about massive neutrinos through the suppression of three-dimensional  matter power spectrum in the presence of massive neutrinos. Using Fisher matrix method, 
we take into account both the detector noise and the cosmic variance owing to the signal. Our statistical estimator allows multiple probes of
the underlying three-dimensional power spectra. The most interesting is the one  in which the covariance of the two-point correlation function
of both the spectroscopic noise and the signal could nearly vanish. We investigate the implications of this 'zero noise' situation in detail.  In addition,  we assess the impact of  priors on
  both cosmological parameters from  CMB data and parameters that determine the thermal and dynamical
 state of the clouds on the signal-to-noise of the detection of neutrino mass.   Finally,  we carry out principal component analysis 
 of the Fisher matrix to understand  the degeneracy  of the 
 extracted quantity (the neutrino mass) with  other  parameters. 

 In the next section, we briefly summarize the semi-analytic method and derive the one-dimensional correlation function.
 The covariance of the correlation function for both the signal and the noise is derived in Appendix~\ref{app:A}. Next we introduce the Fisher matrix for our observational setting, explain the choice of parameters, and present our main results. 
 The final section is reserved for the summary 
 of our main results and concluding remarks. 
Throughout this paper, we use the bestfit parameters of the spatially flat model as estimated with the Planck CMB data \cite{aghanim2020}.

\section{Correlating Lyman-$\alpha$ observations}
\label{sec:sec1}

Lyman-$\alpha$ clouds are absorption features observed  along the line of sights to high-redshift quasars 
(blueward of the Lyman-$\alpha$ 
peak in quasar spectra). These absorption features are caused by neutral hydrogen in regions of mildly non-linear 
baryonic density contrast (e.g. \cite{rauch1998lyman}). These observations measure the flux decrement in the observed quasar spectrum 
$\delta_{\rm F}$. We further define: $\delta_{\rm F} = 
(F-\bar{F})/\bar{F}$, where $F$ and $\bar{F}$ are  the observed and continuum fluxes of the observed quasar. 
$\delta_{\rm F}$ is a function of distance along the line of sight and hence, can be expressed as a function of redshift.  
The Lyman-$\alpha$ clouds have 
HI column densities in the range $N_{\rm HI} \simeq 5\times 10^{12}\hbox{--}10^{15} \, \rm cm^{-2}$ and 
temperature $T \simeq 10^4 \, \rm K$.  These clouds become optically thick   to Lyman-$\alpha$ photons 
for $N_{\rm HI} \gtrsim  10^{14} \, \rm cm^{-2}$. The distribution of HI in Lyman-$\alpha$ clouds
 trace baryonic density fluctuations. Under the assumption that the baryon number density $n_b(z)$
at a given redshift $z$ follows 
a lognormal distribution \cite{coles1991,bi1997}, the 
baryon number density bears the following relation 
with the baryon density contrast $\delta_b(z)$:
\begin{equation}
n_b(z) = n_0(z) \, \,  {\rm exp} \left( \delta_b - \langle \delta^2_b \rangle/2  \right) 
\label{eq:nb}
\end{equation}
where $\delta_b(z)$ follows the Gaussian distribution
and $n_0(z)  = \langle n_b(z) \rangle$ gives the 
mean baryon number density of the IGM at redshift $z$.

At any given redshift, 
\begin{eqnarray}
\delta_F (z) &=& \left[{\rm exp} (-\tau) - 1 \right] \nonumber \\
&=& \sum_{i = 1}^{\infty} \frac{(-1)^i}{i!} \, \, \tau^i.
\label{eq:deltaf}
\end{eqnarray}
Here $\tau(z)$ is the optical depth of Lyman-$\alpha$ clouds. 
From Lyman-$\alpha$ data and
their calibration against hydrodynamical simulations  that 
has been used to model the thermal and dynamical state of 
the clouds, 
it is possible to write $\tau(z)$ as,  
\begin{equation}
\tau(z) = A(z) \, \left( \frac{n_b}{n_0} \right)^{\beta}
\label{eq:tau},
\end{equation}
where $\beta = 2-0.7 (\gamma - 1)$; $\gamma$ is 
the semi-adiabatic index for the Lyman-$\alpha$ clouds that 
has  values in the range $1 \leq \gamma \leq 1.6$,   
and $A(z)$ can be expressed as,  
\cite{hui1997,choudhury2001,pandey2012,sarkar2021}
\begin{equation}
A(z) = 0.946 \, \left( \frac{1+z}{4} \right)^6 \, \left( \frac{\Omega_b h^2}{0.022} \right)^2 \, 
\left( \frac{T_0}{10^4 {\rm K}} \right)^{-0.7} \, \left( \frac{J}{10^{12} {\rm s}^{-1}} \right)^{-1} \, 
\left( \frac{H(z)}{H_0} \right)^{-1}.
\label{eq:coeff}
\end{equation}
Using  Eqs.~(\ref{eq:tau}) and ~(\ref{eq:nb}), 
it is possible to write the optical depth $\tau(z)$ in terms of the 
the underlying baryon density field $\delta_b(z)$. This formalism allows one 
to treat large values of both the  density contrast $\delta_b(z)$ and the optical 
depth $\tau(z)$. In our work we only consider large density contrast and and only
the linear term in the expansion given by Eq.~(\ref{eq:deltaf}). We note that this means
 only a subset of data such that $\tau <  1$ is included in the analysis. This means either analysing only  the low 
redshift data (Eq.~(\ref{eq:tau}))  and/or excluding  Lyman-$\alpha$ clouds with $N_{\rm HI} 
\gtrsim 10^{14} \, \rm cm^{-2}$ from the analysis \footnote{It should be noted that this is the only way to linearly relate  the observable (flux contrast) to the underlying density field (optical depth) and would be applicable not just for  the semi-analytic approach we adopt here. It is possible to use the transition from $\tau < 1$ to $\tau > 1$ in redshift space to constrain density field of the underlying cosmological model \cite{pandey2012,sarkar2021}.}   The high spectral-resolution Lyman-$\alpha$ data shows that only a small fraction of Lyman-$\alpha$ clouds are optically thick for $z \lesssim 2.5$  (e.g. \cite{day2019power}). This information is also 
captured in the effective optical depth, $\tau_{\rm eff}$: 
$\tau_{\rm eff} \lesssim 0.2$ at $z \lesssim 2.5$ (e.g. \cite{faucher2007direct}).  This shows that only a small fraction of data needs to be excluded for $z \lesssim 2.5$. Our proposed estimator, two-point correlation function in redshift space, is not biased by the  missing data (for details of estimating two-point function for the case of missing data see e.g. \cite{Chatterjee2024,Patwa2021} and references therein).

The assumption that $\tau < 1$ further allows us to express the two-point 
correlation function of the flux contrast (Eq.~(\ref{eq:deltaf})) as: 
\begin{equation}
\mathcal{C}_{FF}(z_a, z_b) \equiv \langle \delta_F(z) \delta_F(z') \rangle =  \langle \, \tau(z_a) \, \tau(z_b) \, \rangle \, = \, A^2 \, \times \, e^{ (\beta^2 - \beta) \xi(0) } \, \times \,  
e^{ \beta^2 \xi_{ab} }
\label{eq:CFF}
\end{equation}
Here $\xi(\Delta z)$ is the 2-point correlation function of $\delta_b(z)$:
 \begin{equation}
\xi(\Delta z) \, = \, \langle \, \delta_b(z) \, \delta_b(z^{'}) \, \rangle,
\end{equation}
 $\Delta z = |z - z^{'}|$ gives the absolute magnitude 
of the separation between  redshifts  $z$ and $z^{'}$, 
and $\xi(0) = \langle \delta_b(z) \delta_b(z) \rangle$, 
gives the RMS of the baryon density field $\delta_b(z)$. 

The two-point correlation function of  $\delta_b(z)$
for a given redshift separation $\Delta z$ is computed using 
one-dimensional baryon power spectrum $P_{1D}(k_{1D}, z)$ as, 
\begin{equation}
\xi(\Delta z) = \frac{1}{2 \pi} \, \int^{\infty}_{-\infty} P_{1D}(k_{1D}, z) 
e^{i k_{1D} r^{'}_z \Delta z} d k_{1D},
\label{eq:xi}
\end{equation}
where $r^{'}_z = d r_z/dz$ 
denotes the 1st  derivative of the comoving distance at the  redshift $z$. 
The one-dimensional baryon power spectrum 
$P_{1D}(k_{1D}, z)$ can be computed using 
the three-dimensional matter power spectrum $P_m (k, z)$  as,

\begin{equation}
P_{1D}(k_{1D}, z) = \frac{1}{2 \pi} \, \int^{\infty}_{|k_{1D}|} \frac{P_m(k^{'},z)}
{\left[ 1 + ( k^{'}/k_J)^2 \right]^2} \, k^{'} dk^{'}.
\label{eq:pkjeans}
\end{equation}
at a  redshift $z$. The three-dimensional power spectrum is obtained 
from the publicly available code CAMB\footnote{lambda.gsfc.nasa.gov/toolbox/camb\_online.html}.

We also  compare the numerical results with 
the fitting formula provided by \cite{eisenstein1999}, that allows one to include massive neutrinos. We find that
the two approaches agree within a few percent for parameters of interest to us. We assume Planck bestfit parameters for our 
study for a spatially flat universe. In Figure~\ref{fig:corrfun}, we show the one-dimensional 
baryonic correction function (Eq.~(\ref{eq:xi})) and the 
the impact of massive neutrinos on the  correlation function. 
The Jeans wavenumber, $k_J$,   can be expressed as: 
\begin{equation}
k_J = \frac{1}{H_0} \left[ \frac{2 \gamma k_B T_0}{3 \mu m_p \Omega_m (1+z)} \right]^{-1/2}.
\label{eq:kj}
\end{equation}
The other symbols used above have their usual meaning. Eq.~(\ref{eq:pkjeans}) accounts for the 
suppression in the baryon power spectrum in redshift range of interest. 
\footnote{ Two assumptions made here need re-stating: 
(a) $P_m(k,z)$ is also the power spectrum
of baryonic gas because baryonic perturbations 
follow dark matter perturbation in 
linear theory for the redshift range of interest, 
(b) we assume the baryonic power is 
suppressed below the Jeans scale $k_J$ at a given 
redshift. However, this suppression 
also depends on the redshift evolution of the gas 
at higher redshifts. If the gas got 
reionized at $z\simeq 7.5$, as Planck results suggest 
\cite{aghanim2020} and the universe 
remained ionized at smaller redshifts, the suppression could be 
different from our assumption but doesn't affect our 
main results. }

The covariance of the flux contrast correlation function 
is also of interest to us here. This function is derived in Appendix~\ref{app:A} 
for both the signal and spectroscopic  noise.

\section{ Fisher matrix } \label{sec:fishmat}

The aim of this paper is to study the feasibility of the detection of massive neutrinos 
using the Lyman-$\alpha$ data. We first discuss the data needed for our purposes. We consider
each quasar line-of-sight to be of length $\Delta z = 0.2$, which corresponds to a comoving length
$r \simeq 200 \, \rm Mpc$ at $z= 2\hbox{--}3$. To reliably estimate the flux contrast, the continuum level
$F_0$ needs to be modelled accurately. This requires the flux contrast to be detected with high  S/N at
finer spectral resolution. 
For Lyman-$\alpha$ clouds such that $\tau < 1$, the absorption
line is unsaturated with a thermal velocity width $\simeq 10 \, \rm km \, sec^{-1}$  (for $T\simeq 10^4 \, \rm K$). 
We assume the spectral resolution to be $\Delta v = 5 \, \rm km \, sec^{-1}$ to avoid line blanketing. 
This corresponds to $\delta z \simeq 5 \times 10^{-5}$ at $z = 2$. 
This yields  nearly $5000$ independent  spectral resolution elements
with length scale range  $0.05\hbox{--}200 \, \rm Mpc$ (Day et~al. \cite{day2019power} present a sample of 87 high-resolution intermediate-redshift QSOs in the redshift range $2 < z < 5$.  The  pixel for their sample corresponds to $\simeq 2.1 \, \rm km \, sec^{-1}$. The S/N per pixel is close to ten and S/N$\simeq 19.5$ per resolution element.  Our envisaged theoretical setting correspond closely to the spectral resolution and S/N of these QSOs).

With this observational setting, we next define the Fisher matrix based on two-point correlation function
along the line of sight. Fisher matrix provides us an estimate of the precision with which it will be possible 
to measure 
the parameters used to model a given observation.  The Fisher matrix $F_{ij}$ for the parameter set 
$\lambda_i$ is computed as (e.g. \cite{sivia2006}):
\begin{equation}
F_{pq} = \sum_{abcd} \, \left[ \mathcal{C}^{FF} (z_a, z_b) \right]_{,p} \, 
\left[ {\rm COV} (z_a, z_b\, ; \, z_c, z_d) \right]^{-1} \, 
\left[ \mathcal{C}^{FF} (z_c, z_d) \right]_{,q},
\label{eq:fisher}
\end{equation}
where $a, b, c, d$  denote indices for individual redshifts i.e. $z_a$. 
The indices $p$ and $q$ denote derivatives with respect to parameters $\lambda_i$.
 The sum in Eq.~(\ref{eq:fisher}) is over all pairs such that 
$\Delta z = |z_a -z_b| = |z_c - z_d|$. 
${\rm COV} (z_a, z_b\, ; \, z_c, z_d)$ 
gives the covariance of the two-point correlation function 
$\mathcal{C}_{FF}(z_a, z_b)$.

The covariance of the two-point correlation function 
${\rm COV} (z_a, z_b\, ; \, z_c, z_d)$ receives contribution from both the signal
and spectroscopic noise and is derived in  appendix~\ref{app:A} (Eqs.~(\ref{eq:COV-EE-appC1}), ~(\ref{eq:COV-appC1}),  and (\ref{eq:COV-NN1}))

\begin{figure}[h]
    \centering
    \psfrag{CDM}[cc][][0.65]{${\rm \Lambda CDM}$ \quad \, \, \, \,  }
    \psfrag{neutrino-mass-5}[cc][][0.65]{\, $0.05 \, {\rm eV}$ }
    \psfrag{neutrino-mass-10}[cc][][0.65]{\, $0.10 \, {\rm eV}$}
    \psfrag{titlex}[cc][][0.85]{$\Delta r \, {\rm Mpc}$}
    \psfrag{titley}[cc][][0.85]{$\xi (\Delta r)$}
    \psfrag{titley1}[cc][][0.85]{$\left( 1 - \xi^{\rm model}/\xi^{\rm CDM} \right) \times 100 \%$}
    \includegraphics[scale = 0.575]{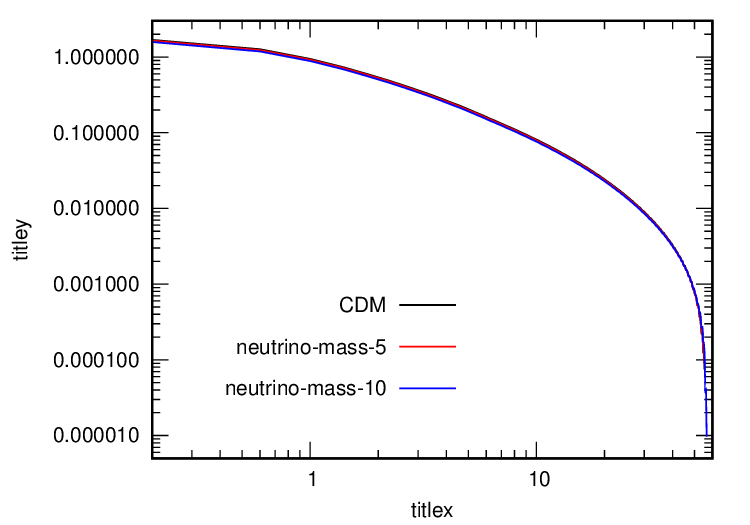}
    \includegraphics[scale = 0.575]{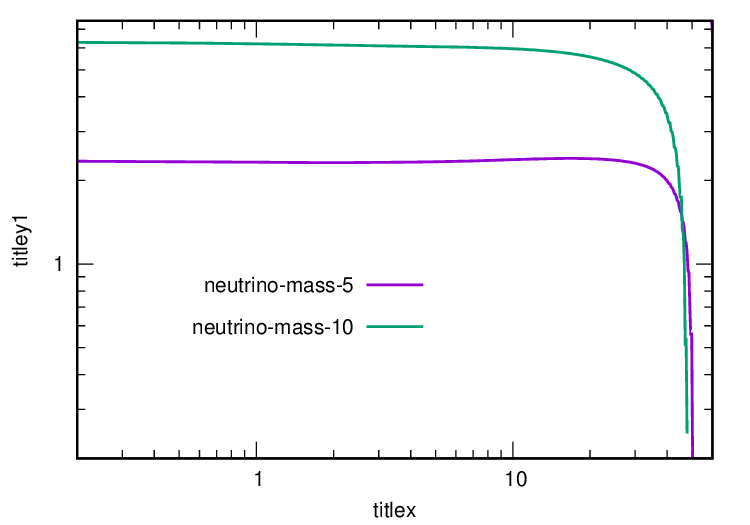}    
    \caption{ Left Panel: One-dimensional correlation function $\xi(\Delta r)$ 
    (see Eq.~(\ref{eq:xi})) is shown as a function of separation $\Delta r$ 
    for the $\Lambda$CDM model and the models with neutrino masses: 
    $m_\nu = 0.05 \, \rm eV$ and $m_\nu = 0.1 \, \rm eV$.  Right Panel: Percentage difference of one-dimensional correlation functions $\xi(\Delta r)$ between  $\Lambda$CDM model and the models with neutrino masses: 
    $m_\nu = 0.05 \, \rm eV$ and $m_\nu = 0.1 \, \rm eV$, 
    as functions of separation $\Delta r$.}
    \label{fig:corrfun}
\end{figure}

In Eqs.~(\ref{eq:COV-appC1}), (\ref{eq:COV-NN1}) and~Eq.~(\ref{eq:COV-EE-appC1}),  the sum is over pairs  ($n_{\rm pair}$) for a  fixed $\Delta z=\Delta z_m$ such that $|z_a - z_b|= |z_c - z_d|=\Delta z_m$. 
The form of the covariance of the two-point correlation  function can be used  to achieve the best sensitivity for the detection of neutrino mass. 
The following choices are possible:
\begin{itemize}
\item[(I)] $a=c$ and  $b=d$. In this case, both the noise and the signal  terms in  Eqs.~(\ref{eq:COV-appC1})  and~(\ref{eq:COV-NN1}) contribute maximally and results
in the worst possible sensitivity. We note that the signal term term dominates the spectroscopic noise term in
this case for $\Delta r \lesssim 2 \, \rm Mpc$. 
\item[(II)] $a\ne c $ and $b\ne d$. In this case, the spectroscopic noise doesn't contribute but the signal 
term's contribution depends on the separation of pairs. 
\item[(III)] $a \ne c $ and $b\ne d$ and widely separated pairs. Under this condition $\xi_{\rm ac}, \xi_{\rm ad}$, $\xi_{\rm bc}$, $\xi_{\rm bd} \ll 1$ (Figure~\ref{fig:corrfun}). This drives both the noise and signal  terms in Eqs.~(\ref{eq:COV-appC1})  and~(\ref{eq:COV-NN1}) to very small values.
If only such pairs are considered in the analysis of data, the covariance matrix becomes nearly singular.   This 
yields a near zero-noise estimator. We consider this case in this paper and find that, for a QSO line-of-sight,    the covariance matrix 
can be nearly independent of $\Delta z$ (for details  see  Figure~\ref{fig:cov1D} and discussion in Appendix~\ref{app:A}). This allows   us to express the covariance as a diagonal matrix:
\begin{equation}
{\rm COV} (z_a, z_b\, ; \, z_c, z_d)  =  \sigma^2_{\rm COV}
\label{eq:sigma_cov}
\end{equation}
where $\sigma_{\rm COV}$ expresses the  effective noise in the measurement of two-point correlation function and encompasses the impact of both the  spectroscopic noise and the  signal covariance. 
\end{itemize}

\subsection{Choice of parameters}
We choose six parameters for our analysis: 
three cosmological parameters:  
$\Omega_{\nu 0} h^2 = \sum m_\nu/(94 \,  \rm eV)$, 
$\Omega_{m0} h^2$, the energy density of cold dark matter, and $n_s$, the scalar spectral index and three 
parameters corresponding to dynamical and thermal state of Lyman-$\alpha$ clouds: $\beta$, $T_0$, and $J_0$.

 The neutrino mass $\sum m_{\nu}$ is given by 
 $\Omega_{\nu 0} h^2 = 
\sum m_{\nu} / 94.22$. 
In addition to $\Omega_{\nu 0} h^2$, we have considered
the following fiducial values of parameters for our analysis: 
$\Omega_{m0} h^2 = 0.14$, 
$n_s =  0.9617$, $\beta = 2.21$ ($\gamma = 0.7$), 
$T_0 = 0.95 \times 10^4 \, \, {\rm K}$, 
$J_0 =1.5 \times 10^{12} \, {\rm s}^{-1}  $. These fiducial values of based on the bestfit values of these parameters
determined from Lyman-$\alpha$ data (for details  see \cite{palanque2020hints}). Our results are insensitive to the choice of these fiducial values. 

While Lyman-$\alpha$ forest  data allows
probe of smaller scales ($k\simeq 5 \, \rm Mpc^{-1}$) that one exploits to determine neutrino masses, this data only
probes a only small range of scales  (typically between 0.5--50~Mpc) and  its interpretation is harder owing to uncertainty in the 
ionization, thermal, and dynamical evolution of the  IGM and Lyman-$\alpha$ clouds (e.g. \cite{palanque2020hints}). As we will see below, the parameter corresponding
to neutrino mass is degenerate with other cosmological parameters such as $n_s$ and $\Omega_m h^2$, which are determined only poorly from the Lyman-$\alpha$ data (for details  see \cite{palanque2020hints}). But these parameters have been determined with great  precision by CMB anisotropy measurements using Planck,  which probe scales from $k \simeq 10^{-4} \hbox{--} 0.1 \, \rm Mpc$ (e.g. \cite{aghanim2020}).
Therefore, any proposal that relies on Lyman-$\alpha$ data to determine neutrino masses gains from the  use of  prior information from CMB 
data.

We  use Planck 1-$\sigma$ error bars as priors for our analysis. We do not choose either the baryon density
$\Omega_b h^2$ or the overall normalization $A_s$ in our analysis as they do not give additional information. Planck determines $A_s$, which normalizes the two-point correlation correlation function,  with high precision but this parameter is degenerate with $\beta$ which is 
not known with comparable precision \cite{palanque2020hints}.  The baryon density  parameter determines the average density which is 
degenerate with $J_0$ (Eq.~(\ref{eq:tau})).

Before embarking on a detailed analysis, we first attempt to establish how well the neutrino mass 
density can be determined in most ideal conditions. If the neutrino mass density is the only parameter
then the error on determining the parameter  from Fisher matrix is $\sigma_\nu = \sqrt{F_{11}^{-1}}$. 
If $\sum m_{\nu} = 0.10 \, {\rm eV}$ ($\Omega_{\nu 0} h^2 \simeq  10^{-3}$),  and  $\sigma^2_{\rm COV} = 10^{-5}$ (as we discuss in  detail in the appendix and Figure~\ref{fig:cov1D}, this value of $\sigma^2_{\rm COV}$ is a modest requirement), 
the error in the measurement of 
$\Omega_{\nu 0} h^2$ is:
$\sigma_{\nu} = 1.4 \times 10^{-4}$.  This gives the SNR of the detection: ${\rm SNR} = \Omega_{\nu 0} h^2 / \sigma_{\nu} \simeq 7.55$ or an approximately 8-$\sigma$ detection. For $\sum m_{\nu} = 0.05 \, \rm eV$, the corresponding  ${\rm SNR} \simeq 3.7$.

The SNR in this ideal condition allows us to benchmark  the degeneracy of the parameter $\Omega_{\nu 0} h^2$ with other parameters. We next include the other cosmological parameters in our analysis. These parameters ($\Omega_{m 0} h^2$ and $n_s$) have been 
measured by PLANCK with high precision. For a spatial flat cosmological model, 
$\Omega_{m 0} h^2 = 0.014240  \, \pm \, 0.00087 (\sigma_{m})$ ; 
$n_s = 0.9665 \, \pm \, 0.0038 (\sigma_s)$ \cite{aghanim2020}. We use Planck  errors 
on these parameters as  priors on our Fisher matrix analysis. This can be achieved by modifying the diagonal 
terms as:  $F^{\rm Prior}_{mm} = F_{mm} + 1/\sigma^{2}_m$ and
$F^{\rm Prior}_{ss} = F_{ss} + 1/\sigma^{2}_s$; 
$F_{mm}$ and $F^{\rm prior}_{mm}$ being the diagonal elements of the 
fisher matrix  for 
the cases of without and with prior on $\Omega_{m 0} h^2$. Similarly, 
$F_{ss}$ and $F^{\rm prior}_{ss}$ are 
the diagonal elements of the fisher matrix for the parameter $n_s$ 
for the cases of without and with prior on $n_s$.  

We invert the fisher matrix 
(including priors on $\Omega_{m 0} h^2$ and $n_s$) to determine the error  on the measurement of $\Omega_{\nu 0} h^2$ 
for  $\Omega_{\nu 0} h^2 = 0.001$. 
This gives us:  $\sigma_{\nu} = \sqrt{ \left[F\right]^{-1}_{\nu \nu} }$, where 
$\left[F\right]^{-1}_{\nu \nu}$ gives the element of the 
inverted fisher matrix for the parameter: $\Omega_{\nu 0} h^2$. 
For  $\sigma^2_{\rm COV} = 10^{-5}$,  the 
estimated error on the measurement of 
$\Omega_{\nu 0} h^2$ is: $\sigma_{\nu} = 1.5 \times 10^{-4}$, 
and this corresponds to 
${\rm SNR} = \Omega_{\nu 0} h^2 / \sigma_{\nu} \simeq 7.2$. For $\sum m_{\nu} = 0.05 \, \rm eV$, the corresponding  ${\rm SNR} \simeq 3.7$.
As SNR in this case is the approximately 
same as the previous case, it shows that Planck
errors  on $\Omega_{m0}h^2$ and $n_s$ are small enough to make the Fisher matrix diagonal-dominated such 
that the degeneracy of these parameters with $\Omega_{\nu 0} h^2$ doesn't 
play a role in its determination. In other words, after the Planck results, the determination of  neutrino energy 
density $\Omega_{\nu 0} h^2$  is nearly independent of other cosmological parameters in a spatially flat model.

We next consider parameters 
related to the thermal and dynamical state of Lyman-$\alpha$ clouds. 
These parameters are:  
$J_0$,  the background photoionization intensity, $T_0$, the mean IGM temperature,  and $\beta$;  $\beta$ is related to the
adiabatic index $\gamma$  via the relation $\beta = 2\hbox{--}0.7 (\gamma \hbox{--}1)$.  These three parameters relate to the ionization and thermal
history of IGM  and the  dynamical evolution of Lyman-$\alpha$ clouds in the redshift range $2 \lesssim z \lesssim 5$ (for a detailed review see e.g. \cite{mcquinn2016evolution} and references therein).  The existing data suggest that the  temperature of IGM is nearly constant for $z > 3.5$:
$T_0 \simeq 10^4 \, \rm K$. For  $z \simeq 3\hbox{--}3.5$,  the temperature the medium is raised by approximately a factor of 1.5 owing to the  reionization of singly-ionized helium  \cite{plante2018helium}. The thermal evolution of the IGM in the entire redshift range of interest to us is  consistent with the
hypothesis that the temperature of optically-thin photoionized  primordial  gas 
is nearly independent of  the magnitude of the ionizing flux with a weak dependence on the  spectral index 
\cite{draine2010physics,mcquinn2016evolution}. The Lyman-$\alpha$ forest data is consistent with   the ionizing intensity  $J_0$ 
remaining constant in  the redshift range $2 < z < 4.2$  \cite{faucher2008flat}. The equation of state parameter $\gamma$ has been   determined  using hydrodynamical simulations and, in this paper, we adopt this parameter in the  range suggested by the  comparison  of these simulations with Lyman-$\alpha$ forest data 
\cite{hui1997,lukic2015lyman}. More details on the acceptable range of these parameters are given in
section~4 of \cite{sarkar2021}.

These parameters are not determined 
as well as the CDM energy density or the scalar spectral index are 
(for details see e.g. \cite{palanque2020hints}). 
For the purposes of our analysis, $T_0$ and $J_0$ are partially degenerate. Both 
$J_0$ and $T_0$ appear in the two-point correlation 
$\mathcal{C}^{FF}$ through the coefficient 
$A^2(z)$ (Eqs.~(\ref{eq:coeff}) and~(\ref{eq:CFF})) as a product 
$J^{-2}_0 T^{-1.4}_0$, which suggests that  the change in 
$\mathcal{C}^{FF}$ due to the change in 
the parameter $J_0$ can be  largely compensated by the  change in the parameter 
$T_0$.  
However, the dependence of the parameter $T_0$ also affects
$\mathcal{C}^{FF} (\Delta z)$ through the Jeans 
wavenumber $k_J$ (Eq.~(\ref{eq:kj})), which partially lifts this degeneracy.

We account for the degeneracies  of the Lyman-$\alpha$ cloud 
modeling parameters with neutrino energy density by carrying out fisher 
matrix analysis with six parameters:
$\Omega_{\nu 0} h^2$, $J_0$, $\Omega_{m 0} h^2$ and $n_s$, 
$T_0$, $\beta$.  As discussed above, we  use Planck priors  on 
$\Omega_{m 0} h^2$ and $n_s$ in the analysis. 
In this case, 
for $\Omega_{\nu 0} h^2 = 0.001$ and 
$\sigma^2_{\rm COV} = 10^{-5}$, we get ${\rm SNR} = 0.4$, 
for a single quasar sightline. 
This is significant deterioration in the prospect of detection. It also clearly identifies  the main obstacle in the  detection of massive neutrinos using Lyman-$\alpha$ forest: the parameters needed to model the Lyman-$\alpha$ clouds
are strongly degenerate with $\Omega_{\nu 0} h^2$.

To understand degeneracy between different parameters, we next consider Principal Component Analysis (PCA) of 
the Fisher matrix. The Fisher matrix (and its inverse) is a symmetric, positive semi-definite matrix. Therefore, it
can a diagonalized by a similarity transform: $F =  O^T W O$. Here $O$ is an orthogonal matrix with unit determinant
and $O_T$ its transpose. $W$ is a diagonal matrix whose  elements  $\lambda_i\ge 0$. The rows of $O$ are the (orthonormal) eigenvectors
of the Fisher matrix. Equivalently, the Fisher matrix can be expressed as outer product of eigenvectors ${\bf A}_i$, weighed by 
the eigenvalues, $\lambda_i$: ${\bf F} = \sum_ i \lambda_i {\bf A}_i\otimes{\bf A}_i$. In this formalism, it is easier to discern both the hierarchy of how well the  parameters are determined and their
degeneracies. If there are no degeneracies between  parameters, the eigenvector is dominated by one parameter and 
the value of the corresponding eigenvalue (after scaling by the value of the parameter) determines how well it is determined.
To illustrate this, we first consider the three-parameter case discussed above. 
In Tables~\ref{tab:table1} and~\ref{tab:table2} we display the 
eigenvalues ($\lambda_i$) and eigenvectors 
of the Fisher matrix (for $\sum m_{\nu } = \{0.05, 0.10 \}\, {\rm eV}$) 
that are scaled by their respective parameter values. The choice of  neutrino masses allows one to distinguish between
 normal and inverted hierarchy as the minimum detectable mass for normal hierarchy is  $m_\nu \simeq 0.05 \, \rm eV$, while it is expected to be nearly two times   in inverted hierarchy. 
We notice that eigenvectors 
are dominated by a single parameter in this case. For a diagonal matrix,   the signal-to-noise for the corresponding parameters ${\rm SNR} \simeq  \sqrt{\lambda_i}$. 
As expected the eigenvector dominated by $\Omega_{\nu 0} h^2$ has the smallest eigenvalue. The extent of degeneracy between a pair of  parameters can be gauged by the projection of the eigenvector
in the direction of the other parameter, weighed by the corresponding eigenvalues.  It is significant if $\lambda_1 a^2 \simeq \lambda_2$
where $a$ is the projection of eigenvector corresponding to eigenvalue $\lambda_1$ in the direction of 
the eigenvector corresponding to eigenvalue $\lambda_2$.  As $\Omega_{\nu 0} h^2$ corresponds
to the smallest eigenvalue, its impact on the determination of other parameters is minimal.  If $n_s$ and $\Omega_{\nu 0} h^2$ are considered together: $a = 0.00366$, 
$\lambda_1 = 3.71 \times 10^5$, and $\lambda_2 = 51.88$. In this case, $a$ is small enough to 
ensure $n_s$ and $\Omega_{\nu 0}$ are nearly independent of each other.  Similar conclusion can be reached for other pairs of 
parameters. As discussed above, the outcome in this 
case is driven by the priors provided by highly-precise Planck determination of $\Omega_{m0} h^2$ and $n_s$.

 We next consider the six-parameter case.  We consider two  neutrino masses ($m_\nu = \{0.05, 0.1\} \, \rm eV$) The eigenvalues and eigenvectors in this case are listed in Tables~\ref{tab:table3}--\ref{tab:table6} for two
 cases of interest ($m_\nu = \{0.05, 0.1\} \, \rm eV$) for $\sigma^2_{\rm COV} = 10^{-6}$ and 
 $\sigma^2_{\rm COV} = 10^{-7}$. The values
 of $\sigma^2_{\rm COV}$ considered are motivated by Figure~\ref{fig:cov1D}. 
 The structure of eigenvectors in the six parameter case is considerably more complicated as compared to the three parameter case.
 To understand degeneracies between different parameters we consider the projection of eigenvectors on other parameters as   discussed above and analyse  the degeneracy of $\Omega_{\nu 0} h^2$ with  other parameters. While this doesn't provide an exhaustive understanding of all the degeneracies that impact the determination of $\Omega_{\nu 0} h^2$ as such degeneracies
 could occur between multiple  parameters, this case,  along with the analytic expression for the two-point correlation function (Eq.~(\ref{eq:CFF})), allows us to recognize the most dominant effects. 
 The following inferences can be drawn from the Tables: (a) the eigenvalue corresponding to the eigenvector dominated 
 by the parameter $\Omega_{\nu 0} h^2$ is much smaller as compared to other parameters \footnote{
 We note that even though the noise covariance matrix is diagonal, it is not possible the scale the results for different $\sigma^2_{\rm COV}$
 because of priors. The prior information is included after assessing the impact of noise on the Lyman-$\alpha$ data.}, (b)
 The second and the third row show the strong degeneracy between $T_0$ and $J_0$ we already discussed above (Eq.~ \ref{eq:CFF}), (c) $\beta$ and $n_s$ are also strongly degenerate and it follows from the analysis of Eq.~(\ref{eq:CFF}), (d)  based on the projection of 
 vectors of higher eigenvalues in the direction of the eigenvector corresponding to $\Omega_{\nu 0} h^2$  as discussed above, it can 
 readily be determined that $\Omega_{\nu 0} h^2$ is strongly degenerate with a combination of $J_0$,  $T_0$, and $\beta$. However, 
 as in the three parameter case, this parameter can determined almost independently of the other cosmological parameter after
 the  Planck priors on those parameters are  used.

  We next  explore the possibility 
 of putting priors on $\beta$, $T_0$, $J_0$ and use them for better determination of the neutrino
 mass.  The current data doesn't allow precise determination of these parameters.
We assume the following RMS errors on the  parameters $T_0$ and $\beta$:  $\sigma_{T} = 0.18$,
 and  $\sigma_{\beta} = -0.07$ (for details  see \cite{palanque2020hints}). This gives only a modest 
 improvement in the SNR of the detection of neutrino mass. 
The reported values of $J_0$ lie within a factor of 2 around $J_0 \simeq 10^{12} \, \rm sec^{-1}$ and it
shows minimal evolution with redshift 
(see e.g. \cite{faucher2008flat}). This means that the current constraints on these parameters do not
allow us to improve the detection prospects of the neutrino mass.  This inference is insensitive to the 
assumed values of these parameters for the Fisher matrix analysis. 

In Table~\ref{tab:table7}, we show the projected SNR for the detection of neutrinos masses. For $\sigma^2_{\rm COV} = 10^{-7}$, the two  neutrino masses  can be 
 determined with ${\rm SNR} \simeq 1.1\hbox{--}2.1$. As we show in Figure~\ref{fig:cov1D}, this 
 is the smallest value of $\sigma^2_{\rm COV}$ achievable along a single QSO line of sight as the total
 comoving distance  is limited to $\simeq 150 \, \rm Mpc$ in this case. This means a marginal 
 detection of neutrino mass might be possible using data from a single line of sight. 
 For achieving a smaller
 $\sigma^2_{\rm COV}$, we need to consider cross-correlating pairs  for larger comoving distances. This can be achieved by using data from multiple QSOs, 
 as discussed in detail in the appendix.  An order of magnitude improvement in  $\sigma^2_{\rm COV}$ allows  the detection of neutrinos with  ${\rm SNR} \simeq 3\hbox{--}6$ for the relevant mass range (Table~\ref{tab:table7}).

We use "zero noise" estimator in our analysis (Eq.~(\ref{eq:COV-NN1})). One limitation of such an estimator in a realistic setting is residual noise. To compute
the residual noise, we simulate the two-point correlation  function ($N_{\rm FF}$) of the  detector noise for one line of sight; we use parameters of the QSO sample studied by \citep{day2019power} for the  simulation.  We find that $N_{\rm FF} \lesssim 10^{-10} \hbox{--}10^{-11}$ for separations/pairs  of interest (see the discussion preceding Eq.~(\ref{eq:COV-NN1}) for detail). This is orders of magnitude smaller than the smallest signal covariance ${\rm COV}_{\rm FF} = 10^{-8}$  (Table~\ref{tab:table7}) we obtain. Therefore, the impact of residual noise is not expected to be 
significant on the estimates of SNR quoted in the paper. 

We next discuss other   sources of   systematic error   in our analysis. One issue is the contamination of Lyman-$\alpha$ forest with metal lines, which reduces the amount of data for analysis (e.g. \citep{day2019power} for detail). Another possible source of error is residual error in  continuum subtraction. As we noted above, the need of high spectral-resolution data  is proposed  partly to deal with this issue. Our other assumptions
relate to the physical state of IGM and Lyman-$\alpha$ clouds. Our choice is based on the success of hydrodynamical simulations 
and semi-analytic models in explaining Lyman-$\alpha$ data (e.g. \cite{mcquinn2016evolution} for detail).  We check that our results are not affected the choice of fiducial models within the range suggested by Lyman-$\alpha$ data and hydrodynamical simulations. We also assume $\tau < 1$ in our analysis. As we note above, this assumption is imperative for relating the density perturbation with flux deficit. This assumption removes some data from Lyman-$\alpha$ forest and renders the sampling inhomogeneous, but doesn't either bias the estimator or alters the robustness of the statistical method (e.g. \cite{Patwa2021}). A more detailed 
investigation of  different challenges posed by these theoretical  assumptions and 
possible observational limitations are beyond the scope of this paper.

\section{Summary and Conclusions} \label{sec:summco}

In this paper, we have studied the feasibility of the detection of neutrino mass ($\Omega_{\nu 0} h^2$)
using high spectral-resolution Lyman-$\alpha$ data. We consider the line-of-sight 
two-point correlation function for our study.
  Our approach adopts  Fisher matrix based on 
semi-analytic modelling of Lyman-$\alpha$ clouds. We assume six parameters for our analysis: three cosmological parameters (neutrino energy density,  $\Omega_{\nu 0} h^2$; CDM energy density, $\Omega_{m0} h^2$; scalar spectral index, $n_s$) and three parameters to model the Lyman-$\alpha$ clouds: background temperature and photon intensity ($T_0$ and $J_0$) along with equation of state of the gas in the clouds, $\gamma$.  The choice of these parameters is justified.  The impact of   precise  prior information on cosmological  parameters from Planck data is also taken into account. 
We  carry out principal component analysis of the Fisher matrix
to assess degeneracies between different  parameters and their impact on the determination of $\Omega_{\nu 0} h^2$.  Using Planck priors, we show that $\Omega_{\nu 0} h^2$ can be determined  nearly independently  of cosmological  parameters
scalar spectral index ($n_s$) and  the energy  density of the CDM component ($\Omega_{m0} h^2$).  
The main stumbling block in the  precise determination of $\Omega_{\nu 0} h^2$ is its degeneracy  with parameters needed to model the thermal and dynamical state of Lyman-$\alpha$ clouds.  

We compute the covariance of line-of-sight two-point correlation function, accounting for the contribution of both the spectroscopic  noise 
and the signal. We  show that it is possible
to envisage a situation ( the signal is dominated by small scales ($\lesssim 10 \, \rm Mpc$) whereas the error on the signal 
comes predominantly from the large scales ($\gtrsim 50 \, \rm Mpc$) ), 
in which the covariance can be driven to very small values  for both
the noise and the signal for a  subset of the data  in a realistic setting for $z\simeq 2\hbox{--}3$ (e.g. the high-spectral resolution data analyzed by Day et~al. \cite{day2019power} could be used for our study).
This  near 'zero noise' estimator 
allows for the detection of the  neutrino mass in the range $m_\nu \simeq 0.05\hbox{--}0.1 \, \rm eV$ 
with  ${\rm SNR} = 1.1\hbox{--}2.1$ for a single line-of-sight. We show that, by novel analysis of 
data, we could increase the ${\rm SNR}$ by using  multiple QSOs. In particular, this might allow us 
to detect the relevant range of neutrino masses with  ${\rm SNR} = 3\hbox{--}6$ (Table~\ref{tab:table7}). 

One  of the outstanding aims of modern cosmology is the detection of massive neutrinos. Current cosmological constraints on massive neutrinos ($\sum m_\nu \lesssim 0.09 \, \rm eV$) are stringent enough to permit
either the minimal normal or the inverted hierarchy models.  Currently, no single data set seems capable of this detection as CMB/Galaxy data are available at large scales whereas
the maximal impact of neutrinos is on the small scales (Figure~\ref{fig:corrfun}). High spectral-resolution Lyman-$\alpha$ data can probe scales suitable for this detection 
but it cannot precisely determine other cosmological parameters, which are degenerate with neutrino mass. One of the possible  approaches  to the detection is  to use  the  prior information from large-scale CMB and/or galaxy data to the small-scale Lyman-$\alpha$ data. We follow this approach in this paper. 
We argue that 
with the novel analysis of the existing Lyman-$\alpha$ data 
proposed here, 
massive neutrinos might be detectable. 

The main takeaways of our analysis are: (a) it is possible to construct a near-zero noise estimator to extract cosmological information from the  Lyman-$\alpha$ data, (b)  one important application of such an estimator is to detect   still-elusive massive neutrinos from these data, (c) such a detection might be possible with a small number of QSOs  with high spectral-resolution (e.g. the sample of \cite{day2019power}).

\begin{table}
\fontsize{9pt}{9pt} \selectfont

\setlength{\tabcolsep}{6pt}
\renewcommand{\arraystretch}{1.5} 

\begin{tabular}{|m{2cm}|m{1.5cm}|m{1.5cm}|m{1.5cm}|}

\multicolumn{4}{c}{ $\sigma^2_{\rm COV} = 10^{-5}$; 
\, Priors used: $\sigma_m = 0.00087$, \, $\sigma_s = 0.0038$ } \\
\hline
Eigenvalue & $\Omega_{m 0} h^2$ & $\Omega_{\nu 0} h^2$ & $n_s$ \\
\hline
57.74               & -0.00275    & 0.9998       & 0.00445  \\
\hline	
$1.36 \times 10^5$  & 0.999	  & 0.00255	 & -0.04378   \\
\hline	
$3.84 \times 10^5$  & -0.0437	  & -0.00457     &  0.999    \\
\hline
\end{tabular} 

\caption{ Table of eigenvalues (first column) and the 
eigenvalues (rest of the columns) of the Fisher matrix 
for the 3-parameter case with $\sum m_{\nu} = 0.05 \, {\rm eV}$. }  

\label{tab:table1}
\end{table}

\begin{table}
\fontsize{9pt}{9pt} \selectfont

\setlength{\tabcolsep}{6pt}
\renewcommand{\arraystretch}{1.5} 

\begin{tabular}{|m{2cm}|m{1.5cm}|m{1.5cm}|m{1.5cm}|}

\multicolumn{4}{c}{ $\sigma^2_{\rm COV} = 10^{-5}$; 
\, Priors used: $\sigma_m = 0.00087$, \, $\sigma_s = 0.0038$ } \\
\hline
Eigenvalue & $\Omega_{m 0} h^2$ & $\Omega_{\nu 0} h^2$ & $n_s$ \\
\hline
51.88                & 0.00201     & 0.9999       & 0.00366   \\
\hline	
$1.35 \times 10^5$   & 0.9995      & 0.0019	  & 0.029383  \\
\hline	
$3.71 \times 10^5$   & -0.029375   & -0.00372	  & 0.9995\\
\hline
\end{tabular} 

\caption{ Table of eigenvalues (first column) and the 
eigenvalues (rest of the columns) of the Fisher matrix 
for the 3-parameter case with $\sum m_{\nu} = 0.1 \, {\rm eV}$. } 

\label{tab:table2}
\end{table}

\begin{table}
\fontsize{9pt}{9pt} \selectfont

\setlength{\tabcolsep}{6pt}
\renewcommand{\arraystretch}{1.5} 

\begin{tabular}{|m{2cm}|m{1.5cm}|m{1.5cm}|m{1.5cm}|m{1.5cm}|m{1.4cm}|m{1.4cm}|}

\multicolumn{7}{c}{ $\sigma^2_{\rm COV} = 10^{-7}$; 
\, Priors used: $\sigma_m = 0.00087$, \, $\sigma_s = 0.0038$, \, 
$\sigma_{\beta} = 0.07$, \, $\sigma_{T} = 0.18$  } \\
\hline
Eigenvalue & $\Omega_{m 0} h^2$ & $\Omega_{\nu 0} h^2$ & $\beta$ & $T_0$ & $J_0$ & $n_s$ \\
\hline
5.76 & 0.000191    & 0.9985       & 0.015882    & 0.015810	& -0.0492    & 0.00118 \\
\hline	
380.62	           & 0.0206	  & -0.0441	& -0.2551       & 0.525	     & -0.8093	 & 0.0402 \\
\hline	
$4 \times 10^4$	   & -0.3043	  & -0.0114	& 0.05207	& -0.7905    & -0.4971	 & 0.1799 \\
\hline	
$6.1 \times 10^4$  & 0.125	  & -0.00869	& -0.331        & -0.1053    & -0.1234   & -0.920 \\
\hline	
$1.8 \times 10^5$  & 0.9269	  & -0.00347	& 0.15          & 0.28       & 0.151     & 0.127 \\
\hline	
$8.57 \times 10^5$ & 0.178	  & 0.0278	& -0.894        & -0.0955    & 0.238     & -0.318 \\
\hline	
\end{tabular} 

\caption{ Table of eigenvalues (first column) and the 
eigenvalues (rest of the columns) of the Fisher matrix 
for the 6-parameter case with $\sum m_{\nu} = 0.05 \, {\rm eV}$. }  

\label{tab:table3}
\end{table}

\begin{table}
\fontsize{9pt}{9pt} \selectfont

\setlength{\tabcolsep}{6pt}
\renewcommand{\arraystretch}{1.5} 

\begin{tabular}{|m{2cm}|m{1.5cm}|m{1.5cm}|m{1.5cm}|m{1.5cm}|m{1.4cm}|m{1.4cm}|}

\multicolumn{7}{c}{ $\sigma^2_{\rm COV} = 10^{-7}$; 
\, Priors used: $\sigma_m = 0.00087$, \, $\sigma_s = 0.0038$, \, 
$\sigma_{\beta} = 0.07$, \, $\sigma_{T} = 0.18$  } \\
\hline
Eigenvalue & $\Omega_{m 0} h^2$ & $\Omega_{\nu 0} h^2$ & $\beta$ & $T_0$ & $J_0$ & $n_s$ \\
\hline
4.69 &	0.00016  & 0.998  & 0.0154  & 0.0246	& -0.0552 & 0.00745 \\
\hline
1503.9	& 0.00411	& -0.0519	& -0.2644       & 0.62	    & -0.736	& 0.0224 \\
\hline	
$3.3 \times 10^4$	& -0.2674	& -0.01357	& 0.0823    & 0.7436    & 0.597	   & -0.1048 \\
\hline	
$5.98 \times 10^4$	& 0.0956	& -0.00982	& 0.32      & -0.0381   & -0.117   & -0.9342 \\
\hline	
$1.74 \times 10^5$	& 0.9455	& -0.00187	& 0.142     & 0.226     & 0.141    & 0.118 \\
\hline	
$6.26\times 10^6$	& 0.158	        & 0.03025	& -0.8945   & -0.0921   & 0.2521   & -0.3187 \\
\hline	
\end{tabular} 

\caption{ Table of eigenvalues (first column) and the 
eigenvalues (rest of the columns) of the Fisher matrix 
for the 6-parameter case with $\sum m_{\nu} = 0.1 \, {\rm eV}$. } 

\label{tab:table4}
\end{table}

\begin{table}
\fontsize{9pt}{9pt} \selectfont

\setlength{\tabcolsep}{6pt}
\renewcommand{\arraystretch}{1.5} 

\begin{tabular}{|m{2cm}|m{1.5cm}|m{1.5cm}|m{1.5cm}|m{1.5cm}|m{1.4cm}|m{1.4cm}|}

\multicolumn{7}{c}{ $\sigma^2_{\rm COV} = 10^{-6}$; 
\, Priors used: $\sigma_m = 0.00087$, \, $\sigma_s = 0.0038$, \, 
$\sigma_{\beta} = 0.07$, \, $\sigma_{T} = 0.18$  } \\
\hline
Eigenvalue & $\Omega_{m 0} h^2$ & $\Omega_{\nu 0} h^2$ & $\beta$ & $T_0$ & $J_0$ & $n_s$ \\
\hline
0.792 & 0.0000252    & 0.9981       & 0.01388    & 0.02034	& -0.0559    & 0.0001487 \\
\hline	
467.39	           & 0.00235	  & -0.052	& -0.269      & 0.533	     & -0.8004	 & -0.00523 \\
\hline	
5408.15	   & 0.0326	  & -0.0125	& -0.0518	& -0.838    & -0.54	 & 0.0107 \\
\hline	
$5.71 \times 10^4$  & -0.0521	  & 0.0102	& -0.341        & -0.027    & 0.0899  & 0.9336 \\
\hline	
$1.33 \times 10^5$  & 0.9768	  & -0.00469	& 0.169          & 0.045       & -0.0246     & 0.1202 \\
\hline	
$8.68 \times 10^5$  & 0.205	  & 0.0274	& -0.8827       & -0.0941    & 0.235     & -0.337 \\
\hline	
\end{tabular} 

\caption{ Table of eigenvalues (first column) and the 
eigenvalues (rest of the columns) of the Fisher matrix 
for the 6-parameter case with $\sum m_{\nu} = 0.05 \, {\rm eV}$. }  

\label{tab:table5}
\end{table}

\begin{table}
\fontsize{9pt}{9pt} \selectfont

\setlength{\tabcolsep}{6pt}
\renewcommand{\arraystretch}{1.5} 

\begin{tabular}{|m{2cm}|m{1.5cm}|m{1.5cm}|m{1.5cm}|m{1.5cm}|m{1.4cm}|m{1.4cm}|}

\multicolumn{7}{c}{ $\sigma^2_{\rm COV} = 10^{-6}$; 
 \, Priors used: $\sigma_m = 0.00087$, \, $\sigma_s = 0.0038$, \, 
$\sigma_{\beta} = 0.07$, \, $\sigma_{T} = 0.18$ } \\
\hline
Eigenvalue & $\Omega_{m 0} h^2$ & $\Omega_{\nu 0} h^2$ & $\beta$ & $T_0$ & $J_0$ & $n_s$ \\
\hline
0.623 &	0.0000205  & 0.9971  & 0.0114  & 0.0339	& -0.0665 & 0.0000972 \\
\hline
229.21	& 0.000617	& -0.0669	& -0.27       & 0.622	  & -0.731	& -0.00351 \\
\hline	
4326.92	& 0.0285	& -0.0141	& -0.0932     & -0.775    & -0.623	   & 0.00773 \\
\hline	
$5.7 \times 10^4$	& -0.0471	& -0.0113    & -0.344      & -0.0292   & 0.097   & 0.932 \\
\hline	
$1.33 \times 10^5$	& 0.979	        & -0.00488    & 0.159    & 0.0385     & -0.0255    & 0.1125 \\
\hline	
$6.37\times 10^5$	& 0.193	        & 0.0297     & -0.879   & -0.0903   & 0.2477   & -0.344 \\
\hline	
\end{tabular} 

\caption{ Table of eigenvalues (first column) and the 
eigenvalues (rest of the columns) of the Fisher matrix 
for the 6-parameter case with $\sum m_{\nu} = 0.1 \, {\rm eV}$. }  

\label{tab:table6}
\end{table}

\begin{table}
\begin{tabular}{|m{4cm}|m{3cm}|m{3cm}|}
\multicolumn{3}{c}{} \\
\hline
$\sigma^2_{\rm COV}$  & $\sum m_{\nu} = 0.05 \, {\rm eV}$  & $\sum m_{\nu} = 0.1 \, {\rm eV}$  \\
\hline
$10^{-6}$ & $0.4$ & $0.8$  \\
\hline
$10^{-7}$ & $1.1$ & $2.1$ \\
\hline
$10^{-8}$ & $3.2$ & $6.2$ \\ 
\hline
\end{tabular}
\caption{ The table shows the projected  SNR for measuring different two neutrino masses considered here  for different  values of $\sigma^2_{\rm COV}$. } 
\label{tab:table7}
\end{table}

\begin{figure*}
    \centering
    \psfrag{ds}[cc][][0.9]{$\Delta r \, {\rm Mpc}$}
    \psfrag{dl}[cc][][0.9]{$\Delta r \, {\rm Mpc}$}
    \psfrag{0.00001}[cc][][0.65]{$10^{-5}$}
    \psfrag{0.000001}[cc][][0.65]{$10^{-6}$}
    \psfrag{0.0000001}[cc][][0.65]{$10^{-7}$}
    \includegraphics[scale = 0.6]{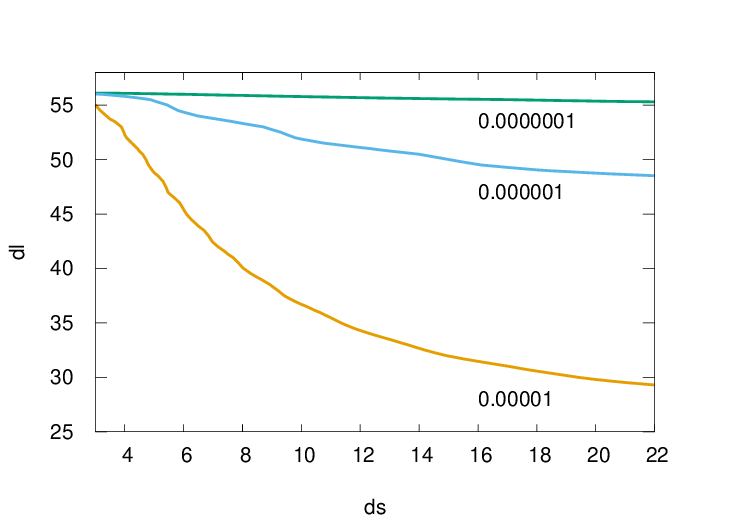}
    \includegraphics[scale = 0.6]{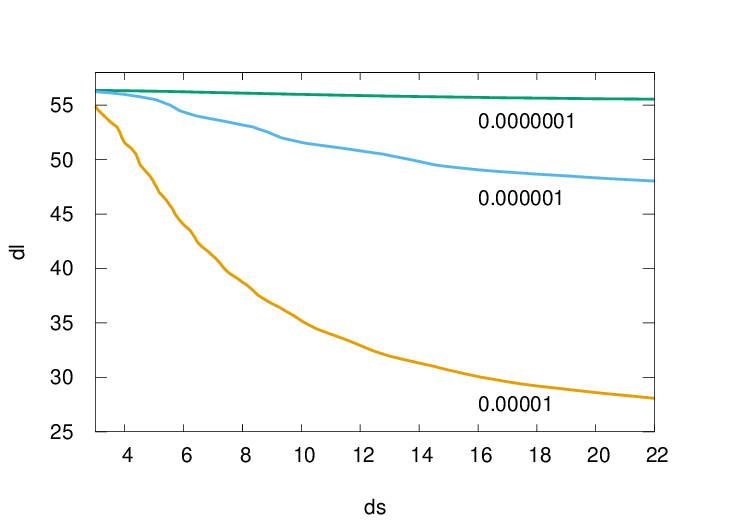}    
    \caption{Contours of fixed 
    $\sigma^2_{\rm COV}$ (Eqs.~(\ref{eq:sigma_cov}) and~(\ref{eq:COV-appC1})) are shown   for $\sum m_{\nu} = 0.05 \, {\rm eV}$ (left panel) and $\sum m _{\nu} = 0.1 \, {\rm eV}$ (right panel).  Each curve is labelled by  a $\sigma^2_{\rm COV}$ value which is computed using 
    Eq.~(\ref{eq:COV-appC1}).}
    \label{fig:cov1D}
\end{figure*}

\appendix
\section{Appendix A: Covariance of the Estimator} \label{app:A}

The starting point of computing the covariance of the two-point function is to define the unbiased estimator
of the two-point function of the flux contrast:
\begin{align}
\hat{E}_{FF}(z_a,z_b) \, = \, \frac{1}{n_{\rm pair}} \, \sum_{a = 1}^{n_{\rm pair}}  \delta_F(z_a) \, \delta_F(z_b)
\label{eq:twopest}
\end{align}
Here $z_b = z_a + \Delta z_{ab}$ and 
 $n_{\rm pair}$ is the number of redshift pairs that correspond
to the same redshift separation  $\Delta z_{ab}$\footnote{The number of pairs could get contribution not only from one line of sight but from multiple QSOs along different 
line of sights}. The flux contrast $\delta_F(z_a)$ receives contribution from
both the signal and spectroscopic noise. Assuming the signal and noise to be uncorrelated, we get:
\begin{align}
\delta_F(z_a) \, \delta_F(z_b) = \hat{C}_{FF}(z_a,z_b) + \hat{N}_{FF}(z_a,z_b)
\end{align}
Here $\hat{C}_{FF}(z_a,z_b)$ and $\hat{N}_{FF}(z_a,z_b)$ correspond to the signal and the spectroscopic noise, 
respectively. 
\begin{equation}
\langle \hat{N}_{FF}(z_a, z_b) \rangle \, = \sigma^2_N \, \delta_{a, b},
\end{equation}
This defines  $\sigma^2_N$, the variance of the spectroscopic   noise. The covariance of 
the  estimator given by  Eq.~(\ref{eq:twopest}) can be expressed as:
\begin{align}
&{\rm COV}(z_a, z_b; z_c, z_d) \, = \, \left \langle \left[ \, \hat{E}_{FF}(z_a, z_b) \, - E_{FF}(z_a, z_b) \, \right] \,  \left[ \, \hat{E}_{FF}(z_c, z_d) \, - E_{FF}(z_c, z_d) \, \right] \right \rangle \nonumber \\ 
\, =& \, \langle \, \hat{C}_{FF}(z_a, z_b) \hat{C}_{FF}(z_c, z_d ) \, \rangle \, - \, C_{FF}(z_a, z_b) \, C_{FF}(z_c, z_d) \, \, \, + \, \, \, \nonumber \\
\, & \langle \, \hat{N}_{FF}(z_a, z_b) \hat{N}_{FF}(z_c, z_d ) \, \rangle \, - \, 
\sigma^4_N \, \delta_{a, b} \, \delta_{c, d}. 
\label{eq:COV-EE-appC1}
\end{align}
where $E_{FF}(z_a, z_b) = \langle \hat{E}_{FF}(z_a, z_b) \rangle$, 
and we have used the fact that the estimator for the Lyman-$\alpha$ 
forest and that for the noise are uncorrelated:
\begin{align}
\left \langle \hat{C}_{FF}(z_a, z_b) \, \hat{N}_{FF}(z_c, z_d) \right \rangle \, =& \, \left \langle \hat{C}_{FF}(z_a, z_b) \right \rangle \, \left \langle \hat{N}_{FF}(z_c, z_d) \right \rangle \nonumber \\ 
=& \, C_{FF}(z_a, z_b) \, \sigma^2_N \, \delta_{c, d}
\end{align}

Eq.~(\ref{eq:COV-EE-appC1}) requires us to compute the expectation value of both the signal and noise four-point functions. It can 
be shown that: 
\begin{align}
&\left \langle \, \hat{C}_{FF}(z_a, z_b) \, \hat{C}_{FF}(z_c, z_d) \, \right \rangle \, = \,
\left \langle \, \delta_b(z_a) \, \delta_b(z_b) \, \delta_b(z_c) \, \delta_b(z_d) \, \right \rangle 
\nonumber \\ 
& \, = \,  
A^4 \times e^{- 2 \beta \xi(0)} \times e^{ \, \frac{\beta^2}{2} \, \left \langle \, ( \, \delta_b(z_a) + \delta_b(z_b) + \delta_b(z_c) \, + \, \delta_b(z_d) \, )^2 \, \right \rangle \, } \nonumber \\ 
&\, = \,
A^4 \, \times \, e^{ 2 (\beta^2 - \beta) \xi(0) } \,
 \times \, e^{\beta^2 \, ( \, \xi_{ab} \, + \, \xi_{cd} \, + \, \xi_{ac} \, + \, \xi_{ad} \, + \, \xi_{bd} \, + \, \xi_{bc} \, ) \, }
\label{eq:4point}
\end{align}
It further follows that: 
\begin{align}
&\left \langle \, \hat{C}_{FF}(z_a, z_b) \, \hat{C}_{FF}(z_c, z_d) \, \right \rangle \, - \,{C}_{FF}(z_a, z_b) \, {C}_{FF}(z_c, z_d)
\nonumber \\ 
&\, = \,  
A^4 \times e^{2 (\beta^2 - \beta) \xi(0) } \times e^{\beta^2 (\xi_{ab} \, + \, \xi_{cd} ) } \times \left[ \, e^{\beta^2 (\xi_{ac} \, + \, \xi_{ad} \, + \, \xi_{bd} \, + \, \xi_{bc} \, ) } \, - \, 1 \, \right]
\label{eq:COV-appC1}
\end{align}
Eq.~(\ref{eq:COV-appC1}) yields the relevant expression for the signal terms in Eq.~(\ref{eq:COV-EE-appC1}). We next consider the noise
terms. It can be shown that: 
\begin{align}
&\frac{1}{n^2_{\rm pair}} \, \left[ \, \sum^{n_{\rm pair}}_{a, c = 1} \, \left \langle \, \hat{N}_{FF}(z_a, z_a + \Delta z_{ab}) \, \hat{N}_{FF}(z_c, z_c + \Delta z_{cd}) \right \rangle \, \right] \, - \, \sigma^4_N \, \delta_{ab} \, \delta_{cd} 
\nonumber \\
& = \, \frac{1}{n^2_{\rm pair}} \, \sum^{n_{\rm pair}}_{a, c = 1} \, \left[ \, \sigma^4_N \, \left( \, \delta_{ac} \, \delta_{bd} \, + \, \delta_{ad} \, \delta_{bc} \, \right) \, \right] \nonumber \\
& = \, \frac{2 \sigma^4_N}{n_{\rm pair}} \, \, \, \, \, \, {\rm for} \, \, \Delta z_{ac}, \Delta z_{ad}, \Delta z_{bc}, \Delta z_{bd} = 0 \nonumber \\
& = \, \frac{\sigma^4_N}{n_{\rm pair}} \, \, \, \, \, \, {\rm for} \, \, \Delta z_{ac}, \, \Delta z_{bd} = 0 \nonumber \\
& = \, 0 \, \, \, \, \, \, \, \, \, \quad {\rm otherwise}, 
\label{eq:COV-NN1}
\end{align}
Eqs.~(\ref{eq:COV-appC1}) and ~(\ref{eq:COV-NN1}) provide  both the signal and the spectroscopic noise  terms in the 
covariance of the two-point function (Eq.~(\ref{eq:COV-EE-appC1})).

It should be noted that the noise
contribution declines as $1/n_{\rm pair}$. $n_{\rm pair}$ can be increased by considering 
 lines of sights to multiple QSOs. The signal contribution in the covariance  (Eq.~(\ref{eq:COV-appC1})) can also be decreased by considering multiple lines of sights but that
 process is more complicated and is discussed below.  
It should be further  noted that if $z_a\ne z_b \ne z_c \ne z_d$ the spectroscopic noise term (Eq.~(\ref{eq:COV-NN1})) is driven to zero.  In other 
words, if two different redshift pairs are used for the Fisher matrix analysis, the covariance might receive contribution
from only the signal covariance. 
We simulate the contribution of the noise 
using the parameters of 
Day et~al. \cite{day2019power} sample, 
and  verify that, for our proposed method of analysis, the covariance of the two-point function is dominated by the signal. It also follows from Eq.~(\ref{eq:COV-appC1}) that if the  pairs are widely separated,
the signal contribution can also be reduced by a large amount as the one-dimensional correlation function for widely separated points
fall very rapidly (Figure~\ref{fig:corrfun}).  In Figure~\ref{fig:cov1D} we show contours of a fixed {\rm COV}$(z_a, z_b; z_c, z_d)$ using Eq.~(\ref{eq:COV-appC1}), in the limit when the separation between the two pairs is much larger than the intra-pair separation.  The x-axis corresponds to the intra-pair separation while the y-axis correspond to
the separation between two pairs. It is clear $\sigma^2_{\rm COV}$ falls sharply for large inter-pair separation and reaches a nearly constant value when the separation between pairs approaches distances 
$\simeq 60 \, {\rm Mpc}$ or so. This motivates us to consider nearly constant $\sigma^2_{\rm COV}$ in the foregoing. We further note that contours shown in   Figure~\ref{fig:cov1D} are suitable for 
a single QSO line of sight of (comoving) length $\simeq 150 \, \rm Mpc$. In such a case, the separation of 
pairs cannot exceed $\simeq 100 \, \rm Mpc$, which puts a lower limit on the value of $\sigma^2_{\rm COV}$. 

However, if we consider multiple QSO lines of sights, $\sigma^2_{\rm COV}$ can be decreased
further. This can be understood  by analyzing the structure of  Eq.~(\ref{eq:COV-appC1}) by considering two QSOs. The two small-scale pairs ($ab$ and $cd$) may correspond to the line of sights of each of the  QSOs 
while the cross-correlation on large scales  ($ac$, $ad$, etc.) is carried out between the different QSOs. Eqs.~(\ref{eq:xi}) and~(\ref{eq:pkjeans}) have
to be suitably modified for this case. Eq.~(\ref{eq:pkjeans}) is derived along a given direction by fixing the other two coordinates  (e.g. \cite{desjacques2004redshift} for details). Since the power spectrum is rotationally invariant, eq.~(\ref{eq:pkjeans})  is left unchanged by rotating the axis such that it connects two points on different QSOs. Similarly,  Eq.~(\ref{eq:xi}) can be cast in terms of the comoving distance between two points rather than just the redshift difference. We note that when the comoving distance is larger than a few hundred Mpcs, it is possible to achieve $\sigma^2_{\rm COV} \simeq 10^{-8}$ (Table~\ref{tab:table7}). We find that simultaneous analysis based  around 10 QSOs in the redshift range
$z \simeq 2\hbox{--}2.5$ could achieve our proposed outcome (e.g. the sample studied by \cite{day2019power}).

\section*{Acknowledgements}

AKS would like to acknowledge the financial aid  and computational resources 
provided by the National Centre for 
Radio Astrophysics, Tata Institute of Fundamental Research, Pune under their regular 
institute Post - Doctoral Fellow program.

\bibliographystyle{JHEP}
\bibliography{ref}

\end{document}